\documentclass[preprint]{aastex}

\begin{document}
\title{GRB: magnetic fields, cosmic rays, and emission from first principles?} 
\author{Andrei Gruzinov}
\affil{CCPP, Physics Department, New York University, 4 Washington Place, New York, NY 10003}

\begin{abstract}

We describe a scenario for large-scale magnetic field generation and particle acceleration in a collisionless collision of cold plasma clouds. A first-principle (i.e. using particles) numerical simulation of this process might be possible. Our scenario is essentially 3D. We argue that {\it large-scale} magnetic fields are not generated in 2D, even in collisionless plasma. 

We calculate and numerically simulate magnetic field generation by relativistic collisionless Kelvin-Helmholtz instability in 2D. Collisionless tangential discontinuity might be more important than collisionless shock, because tangential discontinuity remains unstable even in the hydro limit, when the shock stabilizes.

\end{abstract}

\section{Introduction}

The ``standard model'' of GRBs and afterglows postulates magnetic field and accelerated particles in rough equipartition with the quasi-thermal bulk of the shocked plasma. Such magnetic fields must be generated in and by the blast wave -- simple compression of the pre-existing field is insufficient by many orders of magnitude.

If the blast wave manages to generate the field somehow, particle acceleration must occur (Fermi 1949) giving the observed emission by a combination of synchrotron and Compton (Meszaros \& Rees 1993), and possibly giving the observed ultra-high-energy cosmic rays (Waxman 1995).

We proposed that kinetic (collisionless) plasma instabilities are important for the magnetic field generation (Gruzinov \& Waxman 1999). But this ``collisionless dynamo'' scenario has a severe problem. Weibel instability generates the fields on small scales, roughly of order plasma skin. The fields cannot survive on small scales, non-linear Landau damping should kill them. We therefore {\it speculated} that the length scale of the field grows somehow (Gruzinov 2001).

Recently, a numerical simulation of a 2D strong relativistic collisionless shock has been reported (Spitkovsky 2008, Keshet et al 2008). The authors see the growth of the large-scale field and particle acceleration. That one can do such a thing from first principles is truly a great achievement. There are just particles plus Maxwell in the numerical simulation -- no assumptions like MHD or postulated scattering of the accelerated particles.

But we argue (\S 2) that {\it large-scale} magnetic fields cannot be generated in 2D, even in collisionless plasma. We believe that the current numerical results are not final -- an even  better and longer simulation must show that the magnetic energy fraction in the shocked plasma tends to zero. The zero magnetic energy fraction means that synchrotron emission will be negligible. This kills the model as a model for what happens in GRBs.

We then describe a scenario, essentially 3D, for the large-scale field generation (\S 3). A first-principle (collisionless) numerical simulation of this scenario might be within reach, and might allow to {\it calculate} the GRB emission and cosmic ray production. It might even turn out that collisionless field generation and particle acceleration in 3D is easier to simulate numerically than in 2D. We also discuss the field generation during the afterglow stage.

We calculate the growth rate of collisionless Kelvin-Helmholtz instability (Appendix) \footnote{The relativistic collisionless Kelvin-Helmholtz instability was calculated in the Russian plasma physics literature in the 60s, but we were unable to find the reference.}. We also numerically simulated the instability in 2D.

\section {No large-scale dynamo in 2D}
Weibel instability generates strong  small-scale magnetic field near the shock transition. The field magnitude is of order equipartition, the length scale is of order skin.  But is there an appreciable field at a large distance, many skins, from the shock? 

Numerical simulations give a positive answer, but we will argue differently. We want to know the magnetic field at about $10^9$ skins (Lorentz factor of the $10^{20}$eV cosmic ray divided by the Lorentz factor of the GRB jet). This is well above the resolution of the numerical simulation, and one has to extrapolate. We argue that current 2D numerical results are not in the asymptotic regime. 

We obviously don't have a real theory for the collisionless shock, all we can offer is a speculation. Yet we find this speculation more convincing than the numerical results. 

Suppose large-scale field has been  generated. Then the plasma will behave essentially as a fluid, with Larmor radius replacing the mean free path. In MHD, there is no dynamo in 2D, by the Zeldovich (1957) antidynamo theorem. 

One can argue that accelerated particles are not described by MHD. But the particles with Larmor radius smaller than the scale of the field cannot play any role, and particles with large Larmor radius do not have enough energy. 

One can also argue that small patches of the field, which were generated by the Weibel instability, can merge into large patches. But dragging of the patches with the same sign of the field (in 2D the magnetic field is a scalar) into one region costs energy. This energy must come from the accelerated particles, because the plasma bulk only feels the field squared as pressure, and does not distinguish the sign of the field. Again accelerated particles cannot have enough energy to drag the patches.

One can then argue that accelerated particles might have enough energy if the spectrum is -2 or shallower. But shallower then -2 is impossible by entropy, and -2 exactly does not seem natural, at least this author sees no reason for it. Besides, numerically, Spitkovsky (2008) finds the slope -2.4. 

The absence of large-scale dynamo in 2D does not mean that the particle acceleration is impossible in 2D. To accelerate particles to arbitrarily high energy you must be able to reflect particles of arbitrarily high energy. This can be done without large-scale equipartition fields. It is sufficient to have a slow enough decay of the field as one moves away from the shock. 

\section {Collisionless large-scale dynamo in 3D}

The mechanism described here is essentially 3D for two reasons. First, hydrodynamic dynamo requires 3 dimensions (Cowling 1933, Zeldovich 1957). Second, even supersonic tangential discontinuities are unstable in 3D (Landau \& Lifshitz  1986).

Consider a collisionless collision of two cold plasma clouds. The clouds may have nontrivial density and velocity fields, and nontrivial shapes, say, degenerating into jets. Collisionless instabilities will generate small-scale magnetic fields and pre-accelerate particles. As small-scale magnetic fields and pre-accelerated particles propagate through the plasma, the large-scale plasma motions become approximately hydrodynamic. In particular, large-scale Kelvin-Helmholtz instability should operate, accompanied by Kolmogorov cascade back to small scales. Near the bottom of the cascade, seed magnetic fields already exist. The standard dynamo mechanism should be able to take these seed fields up the Kolmogorov cascade. The plasma becomes magnetized, further boosting the process of particle acceleration, field generation and Kelvin-Helmholtz instabilities. 

It remains to be seen if this scenario is correct. And even if it is correct, it might be hard to do a numerical simulation using particles. But if one can do such a simulation, the reward might be substantial. First, one will be able to calculate the GRB proper emission from first principles. Also, one will be able to calculate the cosmic ray spectrum. 

As an intermediate step, one can try to get a dynamo from the {\it hydro} simulation of a relativistic supersonic tangential discontinuity. Dynamo by a hydro collision of two cold clouds might also be interesting.

Another intermediate step might be a first principle simulation of relativistic collisionless tangential discontinuity. Such discontinuity is Kelvin-Helmholtz unstable (Appendix). As the instability develops, the plasma might be able to enter the hydro regime, and generate the large-scale magnetic fields by the standard dynamo. If this works, the resulting slope of the cosmic ray spectrum may be the real one. 

We must also discuss a related question of the afterglow emission. It might seem that conventional dynamo cannot operate here, because the blast wave propagating into the interstellar medium is stable, there will be no turbulence and hence no conventional hydro dynamo. The real situation might be different for two reasons. Even though the Blandford-McKee (1976) blast wave is stable, it is not an attractor (Gruzinov 2000). The inhomogeneity of the blast wave might be sufficient to generate turbulence and magnetic fields (Milosavljevic et al 2007). Another possibility is to get turbulence from the inhomogeneities of the interstellar medium (Goodman \& MacFadyen 2007). Both inhomogeneities, of the shock and of the interstellar medium, might actually be generated by the inhomogeneous cosmic ray emission from the earlier stages of the blast wave or the GRB proper.

\acknowledgements

I thank Andrew MacFadyen, Milos Milosavljevic, Anatoly Spitkovsky, Eli Waxman, and Weiqun Zhang for many useful discussions.  

This work was supported by the David and Lucile Packard foundation.

\appendix

\section{Relativistic collisionless Kelvin-Helmholtz instability}
We numerically simulated collisionless tangential discontinuity in a cold 2D plasma with charges of equal mass. The box size was about 100 skins. The relative Lorentz factor was about 10. We found generation of magnetic patches with the field energy of about 10\% of equipartition and size of the patches of about 10 skins. 

Thus, collisionless tangential discontinuity is unstable. It generates seed fields and, in principle, might develop into a large-scale dynamo if simulated in 3D in a big enough box.

We will now derive the differential equation describing linear eigenmodes of a cold collisionless 2D shear flow. For the case of tangential discontinuity we also calculate the growth rate of the instability. 

Consider cold unperturbed relativistic shear flow with velocity and particle density fields:
\begin{equation} 
{\bf v}=\left( 0, V(x),0\right) ,~~~n=n(x).
\end{equation} 
The perturbation is $\propto e^{\lambda t+iky}$, with the electromagnetic field
\begin{equation} 
{\bf E}=\left( E_x(x),E_y(x),0\right) ,~~~{\bf B}=\left( 0,0,B(x)\right).
\end{equation} 
Integrating the linearized equation of motion of the charges (or solving the linearized hydro equations, which is equivalent for the cold collisionless plasma), one gets the Ohm's law
\begin{equation} 
j_x={\Gamma ^2\omega _p^2\over \Lambda }(E_x+VB), ~~~j_y={\lambda \omega _p^2\over \Lambda ^2}E_y-\left( {\Gamma ^2\omega _p^2V\over \Lambda ^2}(E_x+VB)\right) '.
\end{equation} 
Here the speed of light is $1$, $\Gamma \equiv (1-V^2)^{-1/2}$ is the Lorentz factor of the unperturbed flow, $\omega _p^2\equiv {4\pi ne^2\over \Gamma ^3m}$ is the square of the plasma frequency, $\Lambda \equiv \lambda +ikV$, prime is the $x$-derivative.

Maxwell equations with this Ohm's law give the following eigenmode equation:
\begin{equation} 
\left( {\Lambda ^2 + \omega _p^2\over \Lambda ^2(\lambda ^2+k^2+\Gamma ^2\omega _p^2)}E_y'\right) '={\Lambda ^2+\omega _p^2\over \Lambda ^2}E_y.
\end{equation} 

For tangential discontinuity, $V(x)=V~sign(x)$, $n(x)=const$, one gets the growth rate
\begin{equation} 
\lambda ^2={\omega _p^2\over 2}\left( \sqrt{1+8{k^2V^2\over \omega _p^2}}-1-2{k^2V^2\over \omega _p^2}\right) .
\end{equation} 
The eigenmodes with $|kV|<\omega _p$ are unstable. The maximal growth rate $\lambda ={\omega _p\over 2\sqrt{2}}$ is achieved at wavenumber $k={\sqrt{3}\omega _p\over 2\sqrt{2}V}$.

\end{document}